\documentclass{aa}

\usepackage{xspace}
\usepackage{graphicx}
\usepackage{float}

\usepackage{amssymb}
\usepackage{txfonts}
\usepackage{multicol}
\usepackage{subfig}

\usepackage{hyperref}
\usepackage{natbib}
\bibpunct{(}{)}{;}{a}{}{,} 

\usepackage{afterpage}
\usepackage{textcomp}
\usepackage{xcolor}
\usepackage{soul,xcolor}
\usepackage{xtab}

\usepackage{lipsum}

\usepackage{placeins}

\newcommand{\cc}{\ensuremath{{\rm cm}^{-3}}\xspace}
\newcommand{\amm}{\ensuremath{{\rm NH}_3}\xspace}
\newcommand{\tex}{\ensuremath{T_{\rm ex}}\xspace}
\newcommand{\tk}{\ensuremath{T_{\rm K}}\xspace}
\newcommand{\sig}{\ensuremath{\sigma_{\rm{v}}}\xspace}
\newcommand{\vel}{\ensuremath{\rm{v}_{LSR}}\xspace}

\newcommand{\msun}{\ensuremath{M_\odot}\xspace}

\newcommand{\kms}{\ensuremath{{\rm km\,s}^{-1}}\xspace}

\renewcommand\linenumberfont{\color{white}}

\begin{document}

  \title{Asymmetric accretion through a streamer onto the pre-stellar core H-MM1}
  
   \author{
        Spandan Choudhury
          \inst{1}
          \and
          Jongsoo Kim \inst{1}
          \and
          Paola Caselli \inst{2}
          \and
          Chang Won Lee \inst{1,3}
          \and
          Jaime E. Pineda \inst{2}
          }

    \institute{
        Korea Astronomy and Space Science Institute, 776 Daedeok-daero Yuseong-gu, Daejeon 34055, Republic of Korea \\
        \email{spandan@kasi.re.kr}
        \and
        Max-Planck-Institut f\"ur extraterrestrische Physik, Giessenbachstrasse 1, D-85748 Garching, Germany       
        \and
        University of Science and Technology, Korea (UST), 217 Gajeong-ro, Yuseong-gu, Daejeon 34113, Republic of Korea
}
    \date{}

 
  \abstract
  {Dense cores inside molecular clouds are hubs of star formation. 
  Cores have been thought to be isolated from their surrounding cloud. However, this idea is challenged by recent observations of streamers that show evidence of mass flow from outside the core onto the embedded protostar. Multi-component analysis using molecular line observations has also revealed the existence of subsonic material outside the traditional coherent boundary of dense cores.}
  {In this study, we aim to probe the extended subsonic region observed around the pre-stellar core H-MM1 in the L1688 molecular cloud in Ophiuchus using multi-component kinematical analysis of very high-sensitivity \amm data.}
  {We used observations of \amm (1,1) and (2,2) inversion transitions using the Green Bank Telescope (GBT). We then fitted up to two components towards the core and its surrounding molecular cloud. }
  {We detect an extended region of subsonic turbulence in addition to the ambient cloud, which shows supersonic turbulence. This extended subsonic region is approximately 12 times the size of and more than two times as massive as the previously detected subsonic material. 
  The subsonic region is further split into two well-separated, velocity-coherent components, one of which is kinematically and spatially connected to the dense core. The two subsonic components are red- and blue-shifted with respect to the cloud component. We also detect a flow of material onto the dense core from the extended subsonic region via a streamer of length $\approx 0.15$ pc ($\approx30000$ au). }
  {We find that the extended subsonic component kinematically associated with the dense core contains $\approx 27\%$ more mass than the core. This material could be further accreted by the core. The other subsonic component contains a mass similar to that of the core mass, and could be tracing material in the early stage of core formation. The H-MM1 streamer is kinematically similar to the ones observed towards protostellar systems, but is the first instance of such an accretion feature onto a core in its pre-stellar phase. This accretion of chemically fresh material by the pre-stellar core challenges our current understanding of a core evolving with a mass that is unchanged since the time of its formation.}

   \keywords{ ISM: kinematics and dynamics -- ISM: individual objects (H-MM1, L1688, Ophiuchus) -- ISM: molecules -- star: formation}

   \maketitle
%

\section{Introduction}

Stars form in cold dense cores embedded in molecular clouds \citep{jaime_ppvii_fila}. 
Cores are characterised by high densities ($\rm n \ge 10^5\, cm^{-3}$) and low temperatures ($\sim$10 K)
\citep{myers_1983_SF_core,myers-benson_1983_dense_core,caselli_2002_dense_core}. Previous observations have also revealed a sharp transition to coherence from the turbulent ambient cloud at the boundary of dense cores \citep{coh_core_barranco_goodman_1998, pineda2010}. 

In our current understanding of low-mass ($\le 1\msun$) star formation, cores are treated as relatively isolated units that might collapse and form protostellar system(s) \citep{terebey_core_collapse,hc2008_cmf, pelkonen2021_cmf, tsukamoto_2023_ppvii_sf}. In numerical simulations, cores are modelled with a mass that does not change throughout their evolution. Recent observations of streamers show asymmetric accretion of material from the disc envelope \citep[in scales of few hundred au][]{tienhao_2023_streamer, flores_2023_streamer, gupta_2024_streamer, hales_2024_streamer}, within the core \citep[few thousand au][]{Felipe2020_streamer}, and even originating outside the natal dense core \citep[up to 20000 au, e.g.][]{Jaime2020_streamer, teresa2022_streamer, taniguchi_2024_streamer}, onto young protostellar systems. Such accretion has been observed from class II systems \citep{Felipe2020_streamer, garufi_2022_streamer} up to very young, deeply embedded class 0 stars \citep{legou_2019_streamer, Jaime2020_streamer}. These observed streamers show accretion of material by the embedded protostellar system. \citet{kuff_2023_infall} suggest that material can also be accreted from outside the dense core via streamers. 
This challenges the model of isolated core evolution, as accretion from outside the core will result in an increase in the core mass, which is often considered to directly relate to the mass of the star that will eventually form within it. 
However, such asymmetric accretion via streamers has not previously been observed towards a pre-stellar core.

\amm inversion transitions are very useful tracers of the physical properties of structures spanning different physical scales, such as molecular clouds, filaments, and dense cores, as it can be detected in a wide range of densities. Although the critical density of \amm is relatively low \citep[a few times $\rm 10^3\,$\cc for the (1,1) and (2,2) transitions,][]{shirley_2015_crit_dens}, the hyperfine structure of its inversion transitions results in the individual hyperfines remaining optically thin even at high volume densities \citep{caselli_2017_amm, Jaime2022_HMM1}. Fitting multiple hyperfines of \amm\ simultaneously allows for precise constraints on the kinematical information. Moreover, a direct measurement of the gas temperature can be obtained if the \amm\ (2,2) line is also reliably detected \citep[e.g. ][]{GASDR1}.

Using multi-component analysis on \amm observations from the Green Bank Ammonia Survey \citep[GAS, ][]{GASDR1}, we detected gas with subsonic turbulence outside the coherent boundaries of dense cores in the molecular cloud L1688 in Ophiuchus \citep{paper_I}. Our results showed the presence of extended subsonic regions outside the previously determined coherent core boundaries, and suggested that the transition to coherence might be rather gradual, in contrast to previous observations. However, due to the insufficient sensitivities of the existing data, the results in \citet{paper_I} were obtained with averaged spectra towards the cores and concentric shells around each core. Therefore, we lacked the spatial resolution necessary to comment on the distribution, exact extent, and kinematics of the extended subsonic region outside the cores.

In this work, we present results from high-sensitivity observations towards the neighbourhood of the pre-stellar core H-MM1 in L1688, using \amm\ (1,1) and (2,2) transitions with the Green Bank Telescope (GBT). L1688 is a molecular cloud in the Ophiuchus star-forming region, at a distance of 138.4$\pm$2.6~pc \citep[][]{l1688_dist_ortiz-leon}. 
H-MM1 is a starless dense core \citep{johnstone_2004_hmm1, parise_2011_hmm1} in the eastern part of L1688. Previous single-dish observations with \amm \citep{harju_2017, paper_I} have shown a clear transition from supersonic to subsonic turbulence at the boundary of the core. \citet{harju_2017} reported a high degree of deuteration in the interior of H-MM1. More recently, \citet{Jaime2022_HMM1} have shown the first direct observational evidence of \amm depletion towards the centre of this core, at densities of $\sim 10^5\,$cm$^{-3}$.
The sensitivity of these data allows us to reliably fit multiple components to the spectra towards the vicinity of the dense core in the native resolution. We have identified and separated the different components in the line of sight, and studied the kinematics of each component in unprecedented detail. \citet{Jaime2022_HMM1} previously reported a velocity structure similar to streamers, seemingly connecting the H-MM1 core to the ambient cloud. However, they did not perform a multi-component analysis.
In this work, we explore that structure with the help of a multi-component analysis.
In Section \ref{sec_GBT_obs}, we describe the \amm observations used in this work. Section \ref{sec_analy} details the procedure used to fit the data and sort the multiple components. We present our primary results in Section \ref{sec_res}, followed by discussions in Section \ref{sec_disc}. Our main conclusions are listed in Section \ref{sec_conclu}.


\section{Observations}
\label{sec_GBT_obs}

The \amm (1,1) and (2,2) data used in this work were taken with the GBT using the on-the-fly technique. Observations were completed in 19 sessions from November 5, 2021 to January 12, 2022 under the project GBT21B-275 (PI : S. Choudhury). We used the seven-beam K-Band Focal Plane Array (KFPA) as the front-end and the VErsatile GBT Astronomical Spectrometer (VEGAS) back-end. We used mode 20 of the VEGAS configuration, which allows eight separate spectral windows per KFPA beam. Each resultant window has a bandwidth of 23.44 MHz with 4096 spectral channels, providing a spectral resolution of 5.7 kHz ($\sim 0.07\, \kms$ at 23.7 GHz). To maximise the time on source, we used in-band frequency switching with a frequency throw of 4.11 MHz ($\approx\ 50.5\,\kms$) for our observations. The GBT beam at 23.7 GHz is approximately 32$''$, with a main beam efficiency of 0.81.

\begin{figure*}[!ht]  
\centering
\includegraphics[width=\textwidth]{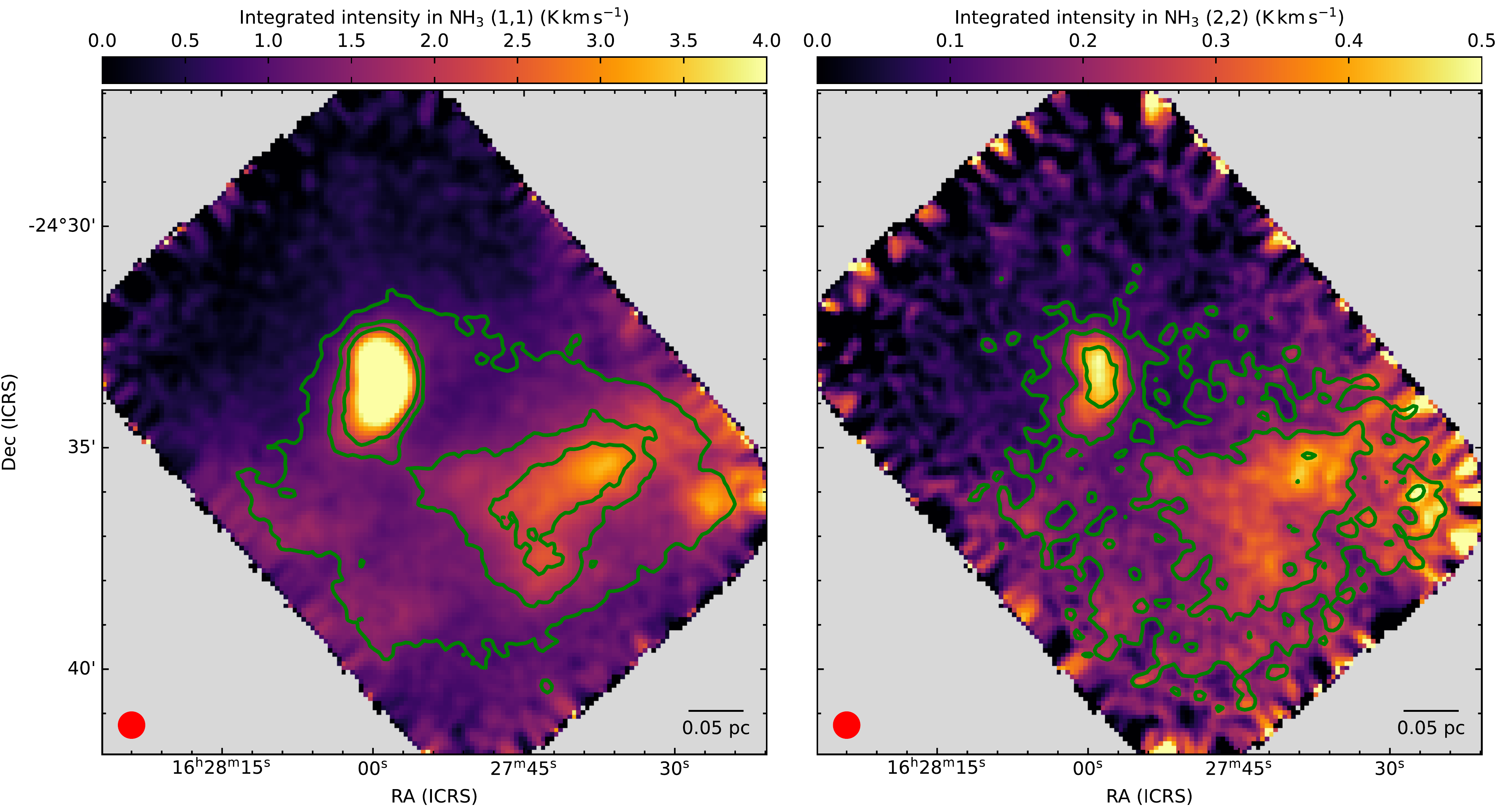}
  \caption{Integrated intensity \amm (1,1) and (2,2). The beam and scale bar are shown in the bottom left and the bottom right corners, respectively. The green contours show 15$\sigma$, 30$\sigma$, and 45$\sigma$ levels for the (1,1) map and 3$\sigma$, 5$\sigma$, and 10$\sigma$ levels for the (2,2) map. The root-mean-square noise levels are 0.08 K \kms and 0.06 K \kms for \amm (1,1) and (2,2), respectively.} 
     \label{fig_mom0} 
\end{figure*}

The observed area was $6' \times 10'$, orientated along the galactic latitude and longitude, respectively, around the pre-stellar core H-MM1. Observations were performed along the galactic longitude to minimise the overhead time. To achieve Nyquist sampling, the separation between each scan row was kept at 13$''$ in galactic latitude. 
For all observations, the scan rate was 4$''\,\rm s^{-1}$, with data dumped every 1.6s. We used a frequency-switching rate of 0.4s to have four samples per integration. The pointing model was updated every 1.5-2 h, depending on weather conditions, using suitable K-band sources.
The quasar 3C 286 was used for flux calibration. For data reduction, we followed the procedure described in \citet{GASDR1}, using \textsc{GBTIDL}\footnote{\href{https://gbtidl.nrao.edu/}{https://gbtidl.nrao.edu/}} \citep{gbtidl_2006}. The final data cubes were generated using all of the completed observations with \textsc{gbtgridder}\footnote{\href{https://github.com/GreenBankObservatory/gbtgridder}{https://github.com/GreenBankObservatory/gbtgridder}}. Figure \ref{fig_mom0} shows the resultant intensity of \amm (1,1) and (2,2), integrated over the entire range of emission. The median noise levels obtained for both cubes were 0.04 K.

\section{Analysis}
\label{sec_analy}

\subsection{Line fitting}
\label{sec_line_fit}

We fitted the \amm\ (1,1) and (2,2) data using the Python package \verb+pyspeckit+ \citep{pyspeckit, pysp_2022}. We used the \verb+cold_ammonia+ model, which is appropriate for regions with temperatures $<40\, $K \citep[see][]{GASDR1}. 
This model produces a synthetic model for all hyperfines of the \amm (1,1) and (2,2) spectra with six independent parameters: kinetic and excitation temperatures (\tk\ and \tex), \amm\ column density (N(\amm)), velocity dispersion (\sig), centroid velocity (\vel), and the ortho-\amm\ fraction of the total \amm column density ($\rm f_{ortho}$). Since we only have detections of the ortho-\amm ((1,1) and (2,2) transitions), we cannot fit for $\rm f_{ortho}$. Therefore, we fixed the ortho-ammonia fraction at 0.5 (assuming an ortho-para ratio of unity) for the fitting process.
Finally, the best-fit model and the corresponding values of the six parameters (with o-\amm fraction fixed at 0.5) were obtained using a non-linear gradient descent algorithm, MPFIT \citep{markwardt2009}. 
Fitting all of the hyperfines simultaneously allows us to obtain very precise constraints on the gas kinematics.
To fit two (or three) components to the data, we added a two-component (or three-component) model as a new model in the fitter, and used this new model for the fit.

We used the Bayesian approach described in \citet{b5_paper} and originally presented in \citet{vlas_2020_bayesian} to fit multiple components and select the optimum number of components in the model. 
The best fit was selected independently for each pixel using the Bayes factor, K.
Following \citet{vlas_2020_bayesian}, we adopted a threshold of $\rm{\ln K^a_b=5}$ to indicate whether model $\mathcal{M}_a$ was preferred over $\mathcal{M}_b$.
$K^a_b$ was defined as
\begin{equation}
\label{eq_k_factor}
    K_b^a = \frac{P(\mathcal{M}_b)Z_a}{P(\mathcal{M}_a)Z_b} ~,
\end{equation}
where $P(\mathcal{M}_l)$s are the posterior probability densities and $Z_l$s are the likelihood integrals over the parameter space. These two parameters were calculated using the likelihood function as 
\begin{equation}
    P(\theta) \propto p(\theta) \times \mathcal{L(\theta)} ~
\end{equation}
and
\begin{equation}
    Z(\theta) = \int_\theta p(\theta) \mathcal{L(\theta)} d\theta ~,
\end{equation}
where $p(\theta)$ are the prior probability densities of the parameters represented by $\theta$.
Finally, the likelihood function, $\mathcal{L(\theta)}$, is related to $\chi^2$ as
\begin{equation}
    \textrm{ln}\ \mathcal{L(\theta)} = \frac{\chi^2}{2} + const.
\end{equation}

As initial priors for the aforementioned parameters (except ortho-\amm fraction), we used suitable ranges to cover the values reported in \cite{paper_I} for these parameters in the different components towards H-MM1 and its neighbourhood. 
The fitting process was iterative and automated. If the fit results indicated that the priors were inadequate for the multi-component fits, their ranges were increased accordingly and the entire process was repeated using the new priors.
The final ranges of priors thus decided for the models were uniformly distributed ranges of $5 - 40\, $K for \tk, $3 - 15\, $K for \tex, $10^{13} - 10^{15}\, \rm cm^{-2}$ for $ N_{\amm}$, $0.06 - 0.8\,$\kms\ for \sig\ , and $2.5 - 5.2\,$\kms\ for \vel. Additionally, in order to avoid having spurious components, the minimum separation between each pair of components was required to be 0.1 \kms\ (approximately twice the spectral resolution), while the maximum separation between two components was set at 2 \kms\ \citep[more than twice the maximum range in velocity across the observed area from single-component fit results, see][]{paper_I} to reduce the parameter space to be covered in the likelihood calculations. \footnote{The scripts and data files used in the analysis presented in this paper, including the scripts to make the parameter maps, can be accessed at \url{https://github.com/SpandanCh/H-MM1_extended_subsonic.}}

\begin{figure*}[!ht]  
\centering
\includegraphics[width=0.51\textwidth]{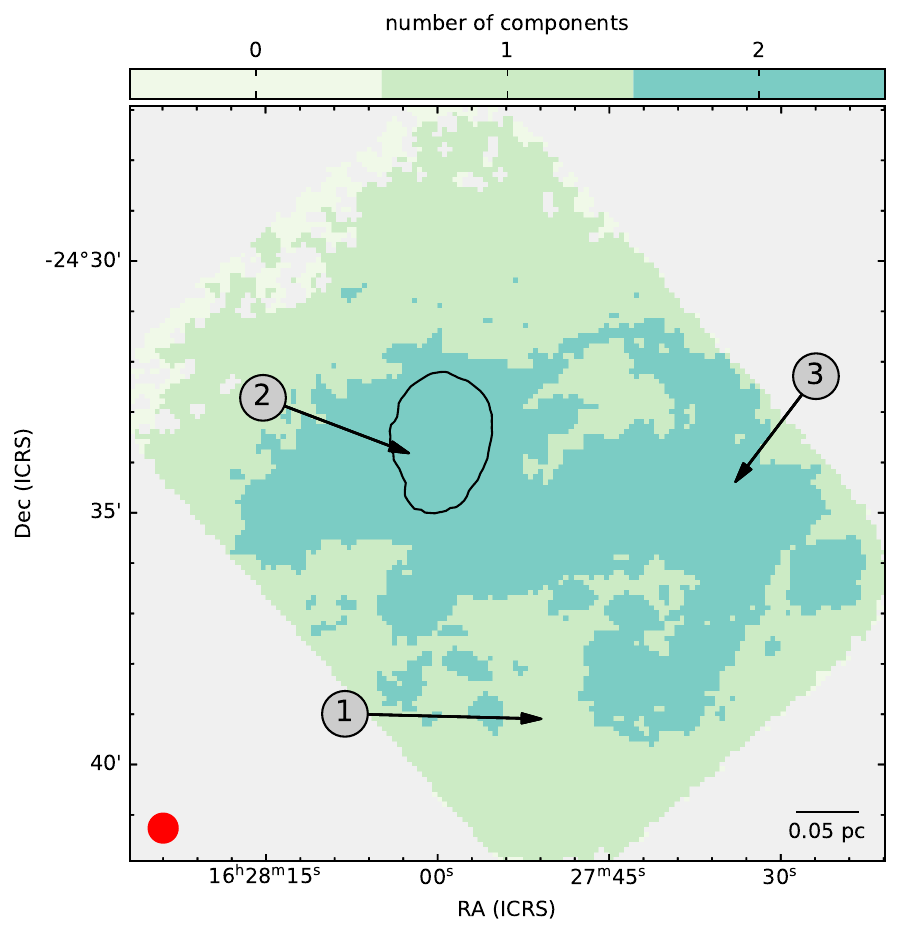}
\hfill
\includegraphics[width=0.46\textwidth]{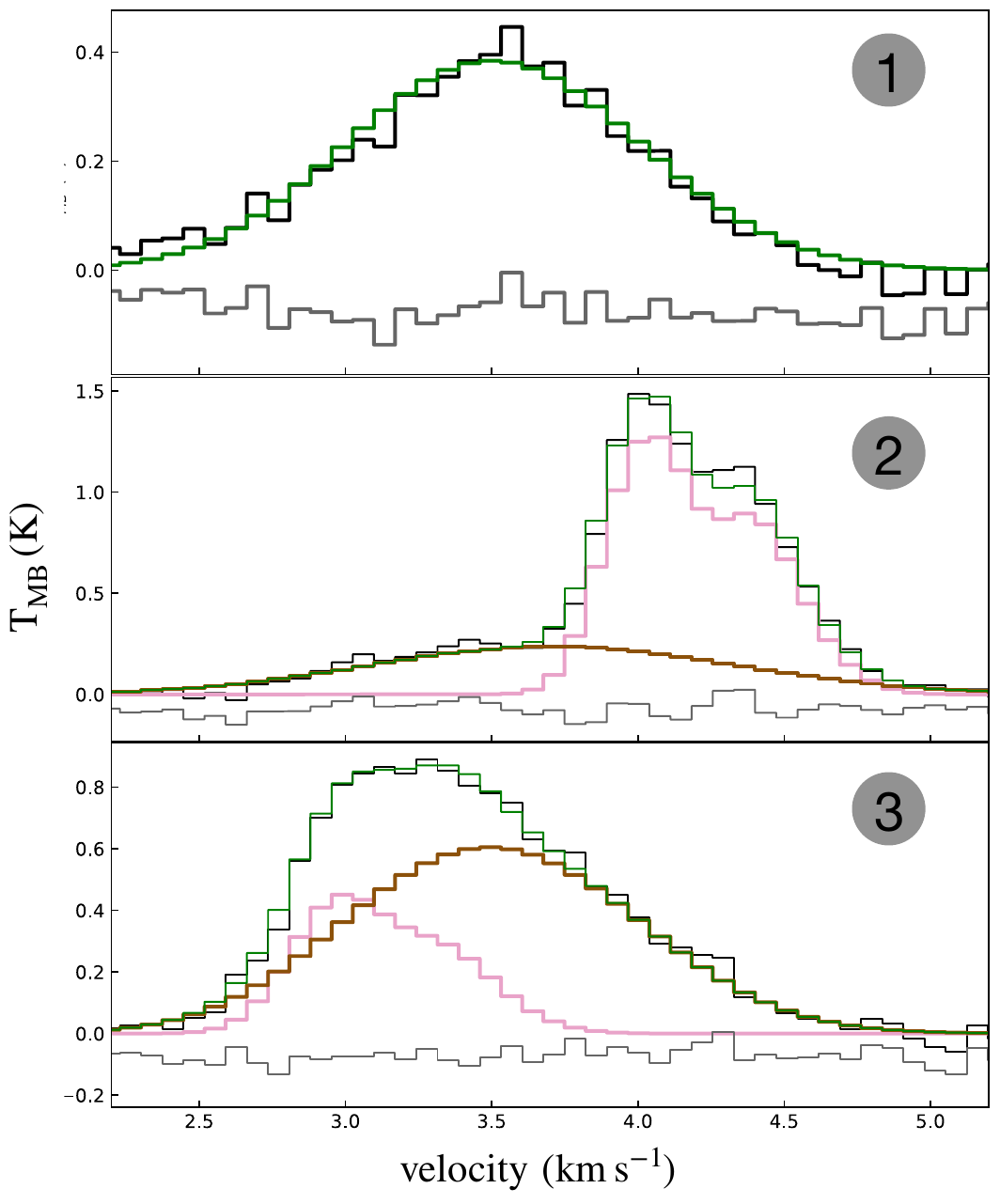}
  \caption{Left: Number of components detected towards the observed region. These components are later grouped into three separate components according to their kinematical properties (see Section \ref{sec_comp_sort}).
  The solid black contour shows the boundary of the coherent region obtained with a single-component fit. The beam and scale bar are shown in the bottom left and the bottom right corners, respectively. 
  Right: Examples of \amm (1,1) spectra with the obtained fit, at the positions indicated in the left panel. Only a small range in velocity is shown here to clearly highlight the individual components.
  In each panel, the fitted model is shown in green, while the residuals are shown in grey. Positions 2 and 3 show two-component fits. The individual components are shown in brown and pink, with the component with the larger linewidth shown in brown.
  Note the multiple, closely spaced hyperfines in the velocity range shown here. The strongest two of these hyperfines can be individually seen in the narrow component (pink) at position 2. For all other cases shown here, these hyperfines are blended together due to the broad linewidth of the spectra.
  }
     \label{fig_num_of_comp}
\end{figure*}

\subsection{Component assignment}
\label{sec_comp_sort}

The left panel of Fig. \ref{fig_num_of_comp} shows the number of components fitted towards different parts of the observed region. 
Apart from some pixels towards the edge, we detect at least one component in the entirety of the observed region. In $\approx$40\% of the pixels, we detect two components in the spectra. 
In the pixels in Fig. \ref{fig_num_of_comp} with a single-component fit (light green), the Bayesian analysis indicates that a one-component model is a better fit than a model without any detection (noise), but a two-component model is not better than a one-component model. Similarly, for the pixels in the bluish-green region, a two-component fit is favoured over a one-component fit, but a three-component fit is not favoured over a two-component fit.
As is seen in Fig. \ref{fig_sig_all_comp}, there is a clear difference between the velocity dispersions of the two components. 
Therefore, we initially considered them as two separate components: `narrow' and `broad'.

The panels on the right of Fig. \ref{fig_num_of_comp} show the spectra at three representative pixels (positions marked by corresponding arrows in the left panel) in different parts of the observed region, along with the model fit. Position 1 shows a pixel with a one-component fit, and positions 2 and 3 show pixels with two-component fits. In all three panels, the total model fit is shown in green, and the residuals (model subtracted from spectrum) are shown in grey. The individual components in the two-component models (for positions 2 and 3) are shown in pink and brown, with the component with the narrower dispersion in pink and the component with the broader dispersion in brown.
For both positions 2 and 3, the velocity of the broad component is similar in both positions ($\approx3.5\,\kms$), while the narrow components at these two positions show clearly different velocities relative to the broad component. This suggests that the narrow component is further subdivided.

Furthermore, the centroid velocity map of the narrow component (left panel in Fig. \ref{fig_vel_all_comp}) reveals that this component is not velocity-coherent. Instead, it clearly has two distinct components with median velocities of 3.23 and 4.09 \kms. Therefore, we further separated the narrow component into two components: narrow-blue and narrow-red, according to their velocity shift from the broad component (Fig. \ref{fig_vel_all_comp}).

\section{Results}
\label{sec_res}

\subsection{Kinematics of the identified velocity components}
\label{sec_kinamatics}

We identified three separate velocity-coherent components with our multi-component analysis. Two of the components (narrow-blue and narrow-red) show subsonic turbulence in the entire region where they are detected, while the other component is turbulent. In our discussion, `turbulence' refers to the non-thermal velocity dispersion, $\sigma_{\textrm{v,NT}}$, calculated by removing the thermal dispersion for the observed molecule (${\sigma_{\rm T}}$) from the total observed velocity dispersion (\sig):
\begin{equation}
    {\sigma_{\textrm{v,NT}}}^2 = {\sig}^2 - {\sigma_{\textrm{T,\amm}}}^2 -{\sigma_{\textrm{chan}}}^2~.
\end{equation}
Here, $\sigma_{\rm chan}$ is the spectral response, which is negligible in this observation \citep[=0.036 \kms,][]{paper_I}. The thermal component of \amm velocity dispersion is 
\begin{equation}
    \sigma_{T,\amm} = \sqrt{\frac{k_B T_K}{\mu_{\amm}}}~,
\end{equation}
where $\rm{k_B}$ is the Boltzmann's constant and \tk is the kinetic temperature in the region, with $\rm{\mu_{\amm}}$ = 17 amu being the mass of the ammonia molecule. 
Then, the sonic Mach number ($\mathcal{M_S}$) is defined as the ratio of the non-thermal velocity dispersion to the sound speed in the medium:
\begin{equation}
    \mathcal{M_S} = \frac{\sigma_{\rm v,NT}}{c_S}~,
\end{equation}
where $c_S$ is the one-dimensional sound speed in the gas, given by:
\begin{equation}
    c_S = \sqrt{\frac{k_B \tk}{\mu_{\rm gas}}}~,
\end{equation}
with ${\mu_{\rm gas}}$ being the average molecular mass \citep[taken to be 2.37 amu, ][]{jens_2008_mu}.

\begin{figure*}[!ht]  
\centering
\includegraphics[width=\textwidth]{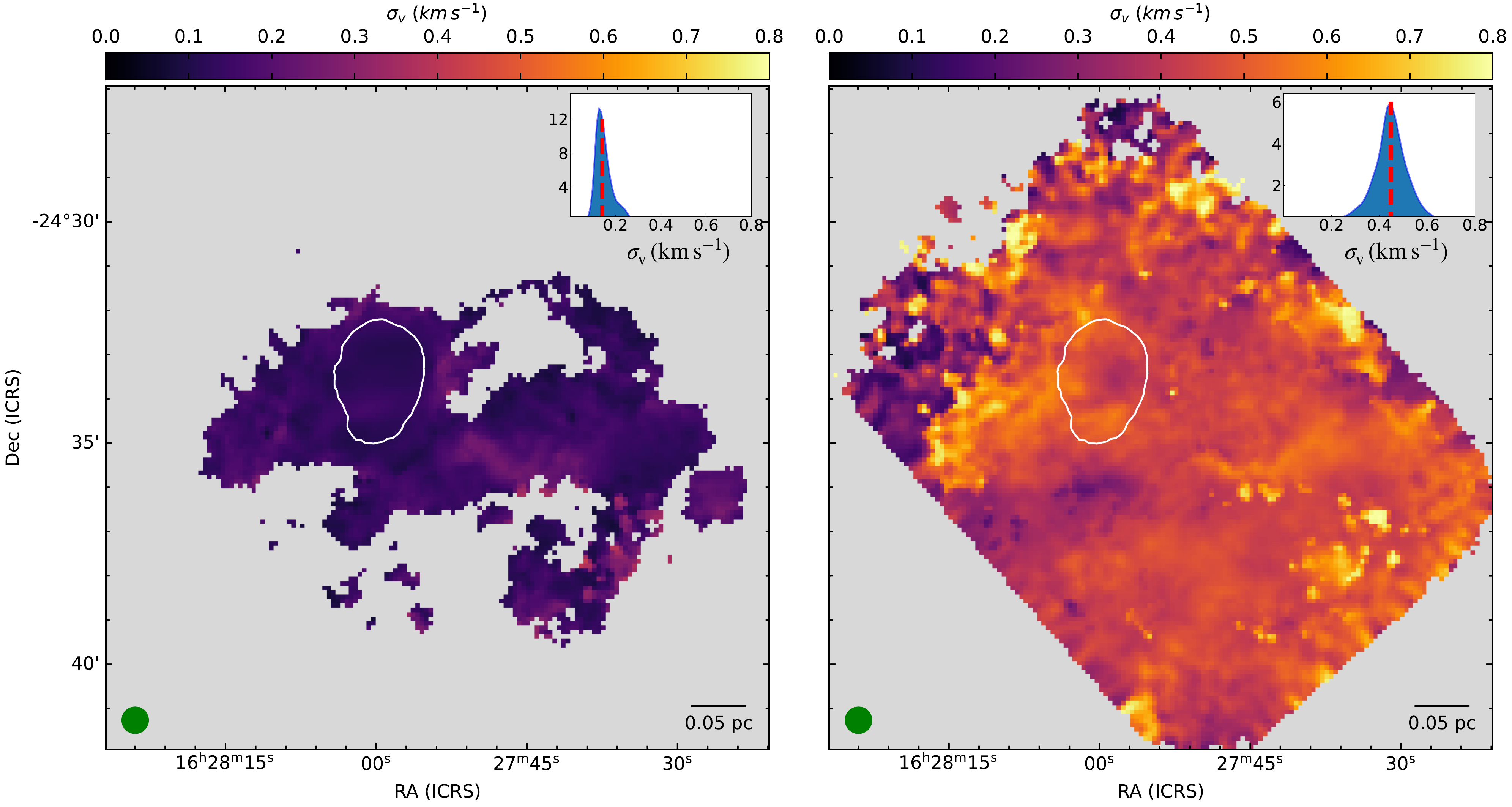}
  \caption{Velocity dispersions of the narrow component (left) and the broad component (right). 
   The probability distribution function of the velocity dispersion of the respective component is shown in the inset in the top right corner of each figure. The vertical red lines in the insets show the median dispersion of each component.
   The turbulence in the narrow component is consistently subsonic throughout, whereas the broad component is always supersonic (the velocity dispersion for sonic turbulence at these temperatures is $\approx$ 0.2\kms).
  The solid white contour shows the boundary of the coherent core from single-component fits. The beam and scale bar are shown in the bottom left and bottom right corners, respectively.}
     \label{fig_sig_all_comp} 
\end{figure*}

\begin{figure*}[!ht]  
\centering
\includegraphics[width=\textwidth]{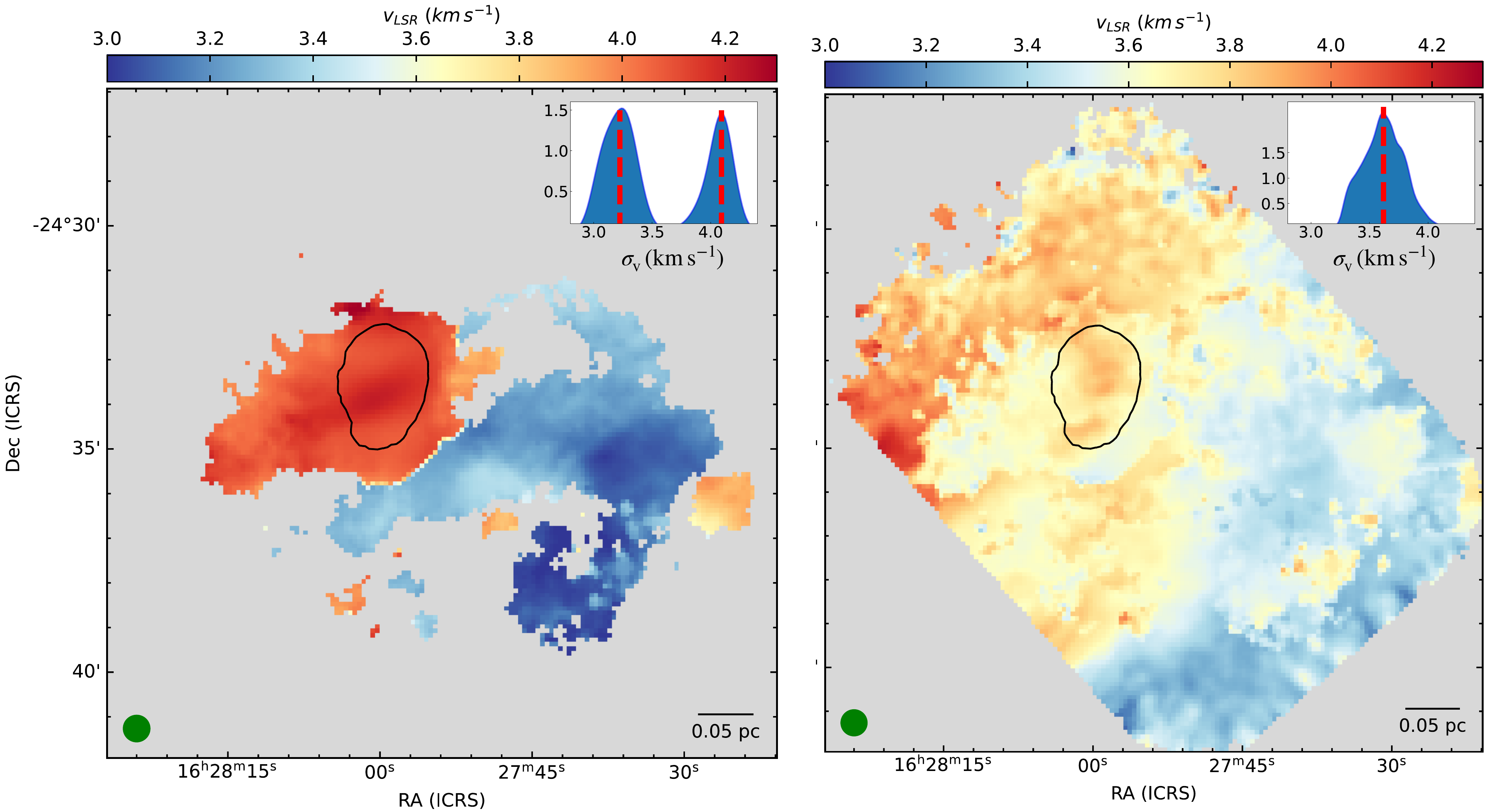}
  \caption{Velocities of the narrow (subsonic turbulence) component (left) and the broad (supersonic turbulence) component (right).
  The probability distribution function of the velocity of each component is shown in the insets in the top right corners of the figures.
  It can be seen that the narrow component is clearly further subdivided into two additional velocity components with median velocities of 3.23 \kms and 4.09 \kms, while the broad component lies at a median velocity of 3.6 \kms. (The median velocities of the different components are shown by the vertical red lines in the insets.)
  The solid black contour shows the boundary of the coherent core from single-component fits. The beam and scale bar are shown in the bottom left and the bottom right corners, respectively. 
  }
     \label{fig_vel_all_comp} 
\end{figure*}

The velocity dispersion and the centroid velocities of the components detected in this work are shown in Figs. \ref{fig_sig_all_comp} and \ref{fig_vel_all_comp}. The velocity of the turbulent (broad) component ranges from 3.3 \kms to 3.9 \kms, with a clearly visible large-scale gradient from north-east to south-west. However, it shows no sharp changes in velocity, suggesting that it is a single, continuous component. This fact, combined with the presence of supersonic turbulence ($\mathcal{M_S}=1.8$, corresponding to a median velocity dispersion of $\rm{0.45^{+0.09}_{-0.09}}\,\kms$) in the component, suggests that it is tracing the ambient cloud towards the region. It should be noted that the errors associated with the median values mentioned here (and in the subsequent section) refer to the difference of the $\rm{16^{th}}$ and the $\rm{84^{th}}$ percentiles of the population from the median value.

The two subsonic components (narrow-red and narrow-blue) are red- and blue-shifted, respectively, relative to the cloud component.
The narrow-red component is spatially and kinematically consistent with the core material. Therefore, we conclude that this component is tracing an extended core region, an area larger than that of the previously identified coherent core (shown with black contours in the figures). The velocity of this component is nearly constant, with a median of 4.09 \kms, and the difference between the $\rm{16^{th}}$ and the $\rm{84^{th}}$ percentiles is less than 0.15 \kms. The turbulence in this component is highly subsonic, with a velocity dispersion of $\rm{0.14^{+0.04}_{-0.2}}\,\kms$ ($\mathcal{M_S}=0.6$). 
The narrow-blue component shows a velocity dispersion similar to that of the narrow-red one, with a slightly larger spread. The median dispersion of this component is $\rm{0.14^{+0.06}_{-0.3}}\,\kms$ ($\mathcal{M_S}=0.5$). However, it lies at a clearly distinct velocity of 3.23 \kms. There is a small observable gradient in velocity from west to east. This is also reflected in the difference between the $\rm{16^{th}}$ and the $\rm{84^{th}}$ percentiles, which is 0.26 \kms. We note that there is a small (<0.5 \kms) difference in the velocity of this component in the north-south and east-west directions, which follows the larger-scale velocity fluctuations seen in the cloud component (right panel in Fig. \ref{fig_vel_all_comp}). There is no clear gradient in the velocity of the narrow-blue component relative to the cloud.

\subsection{Temperatures of the components}
\label{sec_tempe}

\begin{figure*}[!ht]  
\centering
\includegraphics[width=\textwidth]{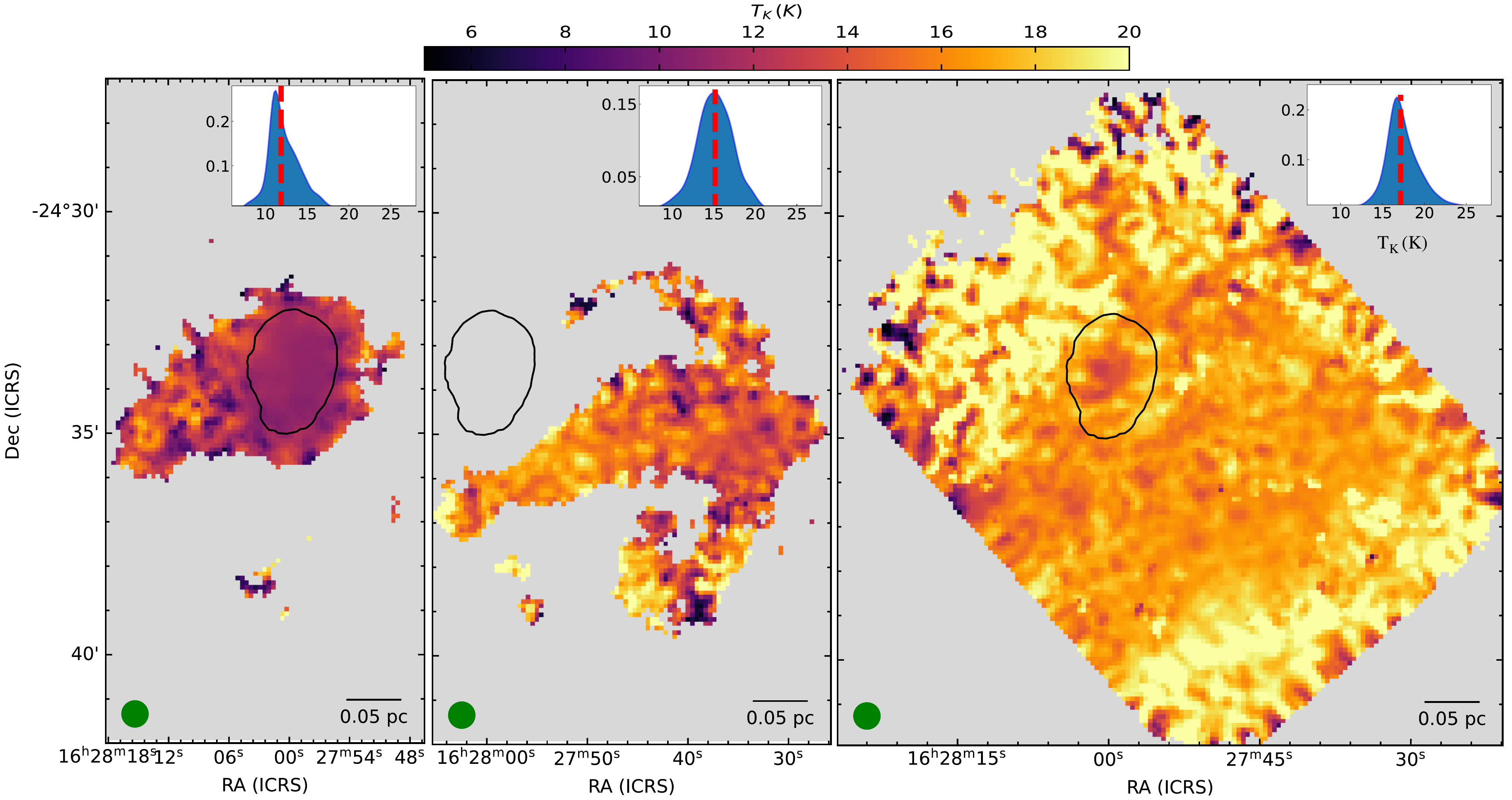}
  \caption{Kinetic temperature of the narrow-red (left), narrow-blue (middle), and broad (right) components. The common colour scale is shown at the top.
  The inset in the top right corner of each figure shows the probability distribution function of the temperature of the respective component. The vertical red lines in the insets show the median of that distribution.
  The narrow component is considerably cooler than the broad component throughout, especially in the vicinity of the coherent dense core, shown by the black contour. The blue-shifted part of the narrow component shows a slightly warmer temperature than the red-shifted counterpart.
  The solid black contour shows the boundary of the coherent core from single-component fits. The beam and scale bar are shown in the bottom left and the bottom right corners, respectively.
  }
     \label{fig_tk_all_comp} 
\end{figure*}

Figure \ref{fig_tk_all_comp} shows the kinetic temperature of the three components, derived from fitting \amm (1,1) and (2,2) lines. We can clearly see a systematic difference in the temperature of the broad component compared to the narrow components. The broad component is at a median temperature of $\rm{17.1^{+2.9}_{-1.7}}$ K, whereas the narrow-red and narrow-blue components are at $\rm{12.0^{+2.8}_{-1.4}}$ K and $\rm{15.2^{+2.6}_{-2.6}}$ K, respectively. 
The difference between the temperature distribution of each component can also be seen from their probability distribution, shown in insets in Fig. \ref{fig_tk_all_comp}.

As was expected, the narrow-red component is at low temperatures characteristic of core material. Similarly, the broad component shows temperatures of around 20 K, which is typical for the warm ambient molecular cloud. On the other hand, the narrow-blue material is at a temperature of ~15 K, which is warmer than typical core material, but cooler than ambient cloud. The east part of this component is at a slightly higher temperature compared to the west, with a difference of approximately 2K.


\section{Discussion}
\label{sec_disc}

\subsection{Extended subsonic region associated with the dense core}
\label{sec_extended_core}

\begin{figure}[!ht]  
\centering
\includegraphics[width=0.48\textwidth]{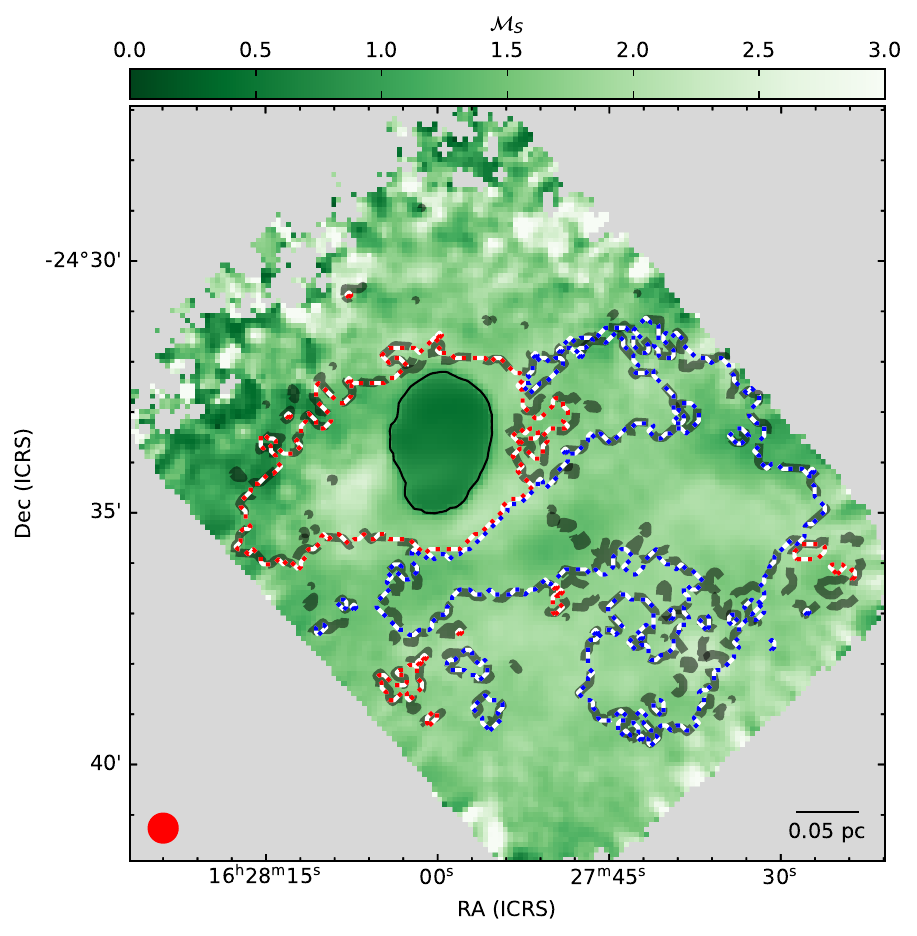}
  \caption{Sonic Mach number, calculated using a single-component fit. The solid black contour shows the boundary of the subsonic region in this map, which is similar to the previously determined boundary of the coherent dense core. The dashed black contour shows the boundary of the extended subsonic region that can only be detected using multi-component analysis as described in this work. 
  The dotted red and dotted blue contours show the extents of the narrow-red and narrow-blue components, respectively.
  The beam and scale bar are shown in the bottom left and the bottom right corners, respectively.}
     \label{fig_mach_comp} 
\end{figure}

Figure \ref{fig_mach_comp} shows the sonic Mach number in the region using a single-component fit to the data. The region with $\mathcal{M_S} \leq 1 $ is shown in solid black contours. As previous studies typically employed a single-component analysis, this contour represents the boundary of the coherent core, determined previously. However, as was mentioned above, with the high sensitivity of our data, and using a multi-component analysis, we are able to detect a much larger region of subsonic turbulence, divided into two velocity-coherent regions: narrow-red and narrow-blue. The combined subsonic boundary of these two regions is shown with dashed black contours in Figure \ref{fig_mach_comp}. With a two-component analysis, we recover a subsonic area that is approximately 12 times larger than the area of the coherent core from previous observations. This previously undetected and substantially larger region of subsonic turbulence indicates that there is much more material available for accretion by the core than previous estimates suggested.

As \amm is a poor indicator of densities, we cannot obtain an accurate measurement of the mass contained in these different regions. However, \citet{anika_2021} showed a correlation between the observed \amm flux and that of dust continuum, yielding a direct conversion between \amm flux and mass. Therefore, we can use the total \amm flux in the respective components in each region as a proxy for the enclosed mass.
We computed the total flux in \amm (1,1) inside the coherent core boundary (solid black contour) in Fig. \ref{fig_mach_comp} using a single-component fit model. This flux can be considered as indicative of the mass of the previously estimated coherent region, as the previous studies used a single-component fit in their analysis. 
Similarly, we can estimate the mass of the extended subsonic region we detect in this work (dashed black contours in Figure \ref{fig_mach_comp}) from the total \amm (1,1) flux in the narrow component (left panel of Fig. \ref{fig_sig_all_comp}). 
Comparing the two, we find that the mass enclosed in the extended subsonic region is 113\% more than that of the previously estimated coherent region.
This means that a significantly large amount of subsonic material was previously undetected due to insufficient sensitivities. This subsonic material could be further accreted by the core (for narrow-red component), or form another core (narrow-blue component).

The narrow-red component, which is kinematically and spatially connected to the dense core, contains additional mass outside the previously calculated core boundary (left panel of Figure \ref{fig_vel_all_comp}). 
Comparing the total \amm (1,1) flux in the narrow-red component inside the core boundary and outside of it, we find that
the amount of additional mass is $\sim$27\% of the core mass. This implies that the mass of the core can increase by this amount during its evolution by accreting the subsonic material that is physically connected to the core. 
Similarly, by calculating the total flux in \amm (1,1) in the 
narrow-blue component, we find that this component contains a mass similar to that of the core ($\approx 90\%$ of the core mass). This component is kinematically similar to core material, but its temperature is intermediate between that of typical dense cores and ambient cloud. This region could be material in the process of forming a core, with a mass similar to that of H-MM1 (see Section \ref{sec_narrow_blue_orig}).

\subsection{Origin of the narrow-blue component}
\label{sec_narrow_blue_orig}

As was discussed in previous sections, in this work we detect three separate velocity-coherent components towards the neighbourhood of H-MM1. Similar to our previous results in \citet{paper_I}, two of those components correspond to material associated with the dense core and the ambient cloud. However, here we detect a third component in the line of sight, which does not appear to represent either the known core or the cloud. This component (narrow-blue) has a velocity dispersion very similar to that of the core, albeit with a slightly larger spread of $\approx 0.1\, \kms$. However, it has a completely different velocity from the core, and lies on the opposite side of the cloud in the velocity plane. 
These features indicate that, although similarly subsonic as the core, the narrow-blue component is tracing a completely separate region.

\begin{figure}[!ht]  
\centering
\includegraphics[width=0.48\textwidth]{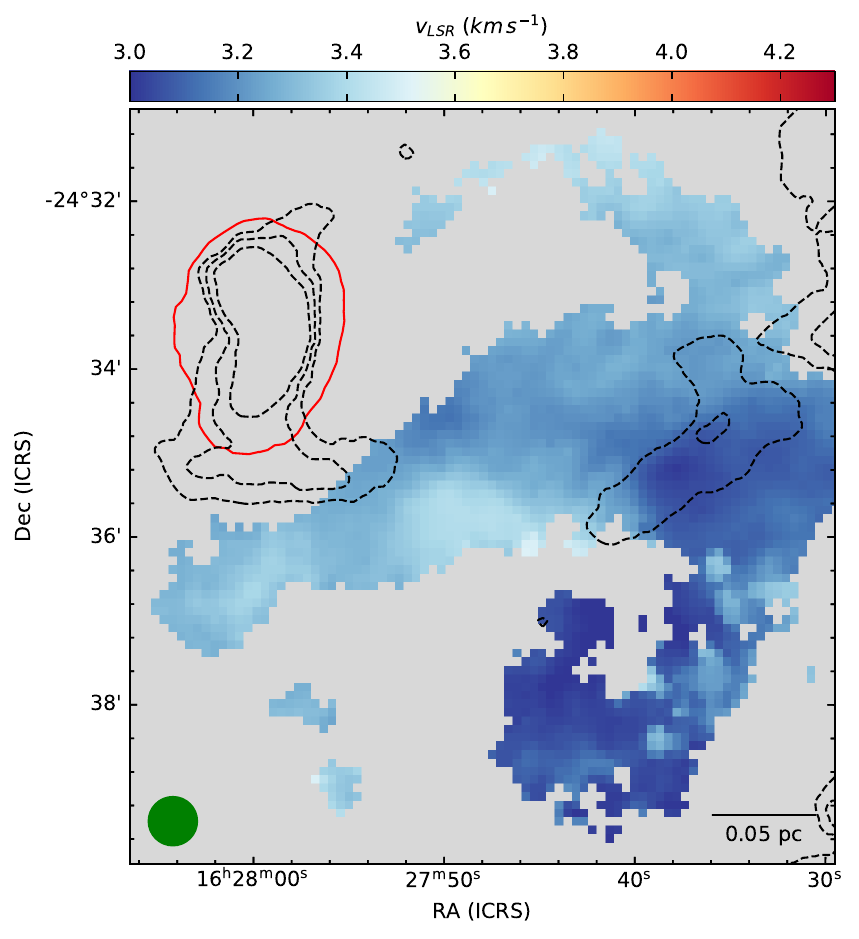}
  \caption{Velocity map of the narrow-blue component, with N(H$_2$) obtained from Herschel observations overlaid as dashed black contours. The contour levels indicate N(H$_2$) $= 1.2 \times 10^{22}, 1.3 \times 10^{22}, 1.5 \times 10^{22}, 1.8 \times 10^{22}\ cm^{-2}$. The red contour shows the boundary of the coherent core from single-component fits.
  The beam and scale bar are shown in the bottom left and the bottom right corners, respectively.
  }
     \label{fig_vel_blue_nh2_contour} 
\end{figure}

This component has a kinetic temperature that is intermediate between characteristic temperatures of typical dense cores and large-scale molecular clouds, which suggests that this region is intermediate between core and cloud. Figure \ref{fig_vel_blue_nh2_contour} shows this component overlaid with contours of N(H$_2$) taken from the Herschel Gould Belt Survey \citep{arzou_2019_fila, ladj_2020_hgbs_oph_strc}. It can be seen that there is a small local peak in the column density around the centre of the region traced by this component. It can be noted that \citet{ladj_2020_hgbs_oph_strc} identify a bound starless core at this location. 

Therefore, the narrow-blue component is tracing gas that has dissipated its turbulence and that is in the process of cooling down from the molecular cloud temperatures. Moreover, it shows a local peak in the column density inside the region.   
These physical and kinetic properties indicate that this component might be tracing material in the early stages of core formation, similar to `Phase II' cores, or coherent, bound cores showing low turbulence, as categorised by \citet{stella_2022_core_evo} using a magnetohydrodynamic simulation of a star-forming cloud.

\subsection{Possible infall into the core}
\label{sec_infall}

\begin{figure*}[!ht]  
\centering
\includegraphics[width=\textwidth]{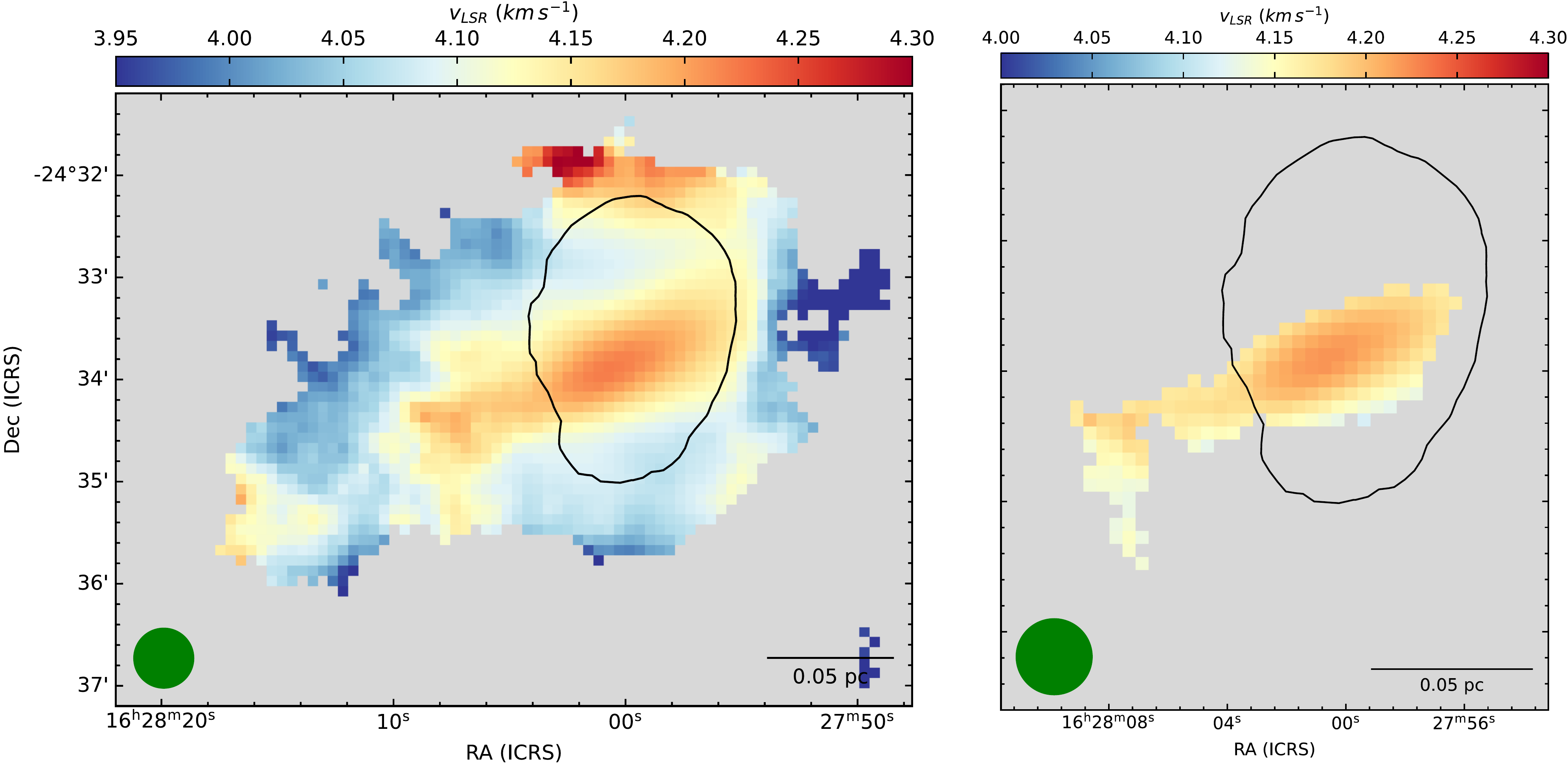}
\\
\includegraphics[width=\textwidth]{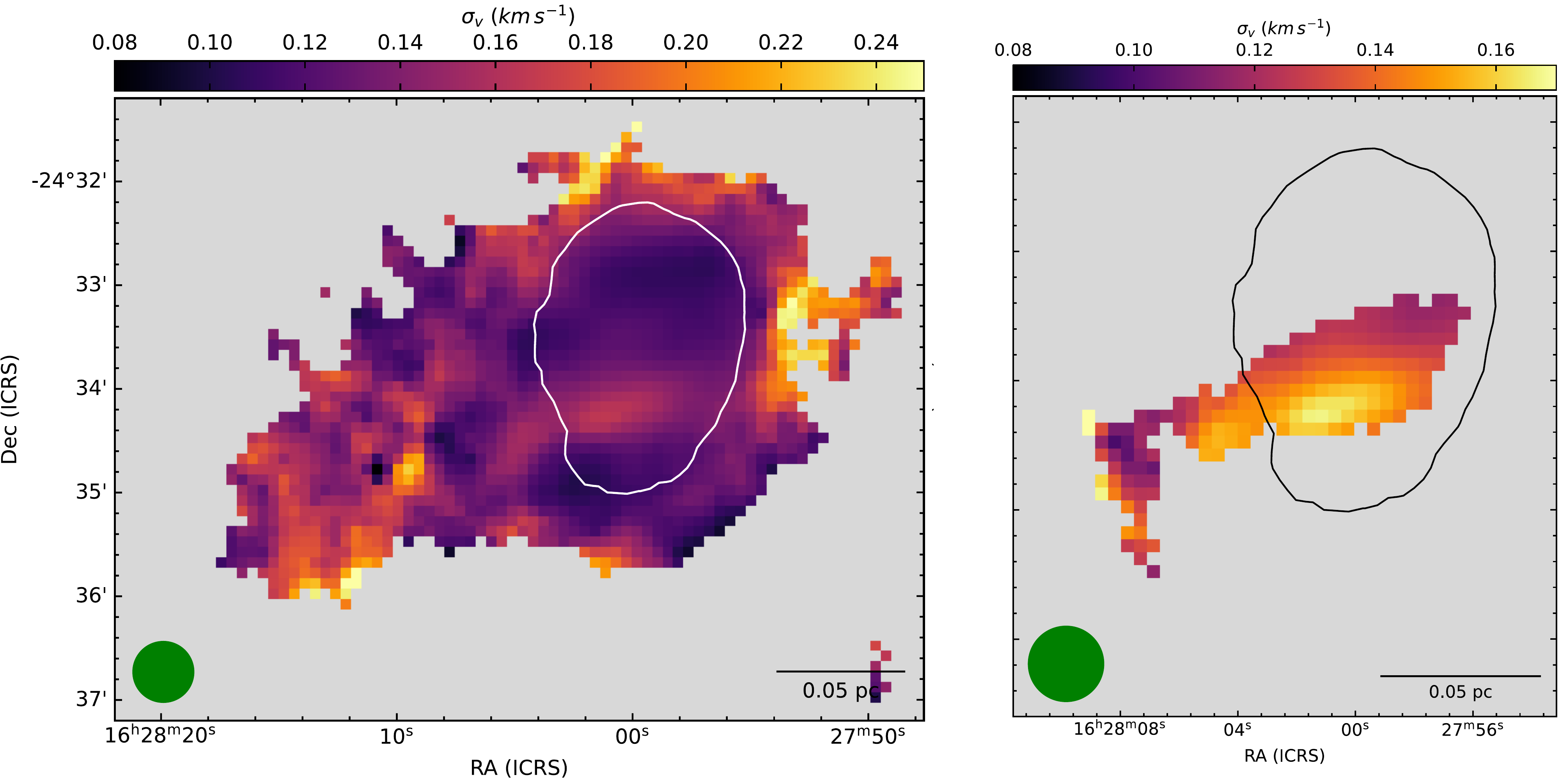}
  \caption{Centroid velocity (top) and velocity dispersion (bottom) of the narrow-red component. The colour scales used here are different from the ones used in the previous figures, with a smaller range chosen to highlight the kinematic features. The panels on the right show the velocity and dispersion of the elongated red-shifted feature.
  The solid black or solid white contour shows the boundary of the coherent core from single-component fits. The beam and scale bar are shown in the bottom left and the bottom right corners, respectively.}
     \label{fig_streamer} 
\end{figure*}

Figure \ref{fig_streamer} shows a zoom-in view of the velocity and velocity dispersion of the narrow-red component. The dynamic ranges shown in this figure are smaller than the ones in the previous figures in order to clearly show an elongated kinematic feature towards the core. This feature was also observed in high-resolution Very Large Array (VLA) observations with \amm towards H-MM1, by \citet{Jaime2022_HMM1}.

This elongated region is slightly red-shifted, at $\approx 4.18\, \kms$, compared to the rest of the core material at $\approx 4.09\, \kms$, and extends from outside the core boundary towards the east, with a total length of $\approx 0.15$ pc, or $\approx 30000$ au. The region where this feature interacts with the core material shows a slightly higher velocity dispersion, with a peak of 0.17 \kms, compared to the rest of the core, where the velocity dispersion is 0.12 \kms. 
Observations of streamers also show a smooth velocity gradient towards the centre, with an increase in the velocity gradient where the streamer meets the protostellar system \citep{Jaime2020_streamer, thieme_2022_streamer, teresa2022_streamer, tienhao_2023_streamer, flores_2023_streamer}. Therefore, the feature observed towards H-MM1 is kinematically similar to streamers. However, it is to be noted that the velocity gradient and the increase in velocity dispersion towards the centre are almost an order of magnitude higher in the case of streamers ($\approx 1\, \kms$ and $\approx 0.5\, \kms$, respectively). This is expected, as the streamers are observed with a much higher resolution, probing very close to the central star, and as there is a protostar at the centre of the infall.

However, in the case of H-MM1, there are no known protostars embedded inside the core. Therefore, it is a much younger system than the ones of the previously observed streamers. Furthermore, the larger GBT beam likely causes smoothening of the small-scale kinematical variations. Therefore, the velocity gradient and the increase in velocity dispersion observed in this feature are quantitatively less than the ones observed in streamers.

This elongated red-shifted feature appears to be tracing mass accretion by the core itself from the extended subsonic region outside its boundary. This could be the first evidence of large-scale asymmetric accretion by a pre-stellar core, in that the accreting material is penetrating the core from one specific direction, similar to the streamers observed towards protostellar discs.

The relatively large GBT beam likely smoothens out the velocity gradient in this structure, which makes further kinematic analysis difficult, especially since the velocity difference between different parts of the observed region is already quite small. Interferometric spectral line observations are required to trace this feature to its largest extent, and compare its kinematics with infall models, similar to streamers. 
Observations with carbon-chain molecules such as CCS, HC$_3$N, and HC$_5$N, which trace chemically young regions \citep{suzuki_1992_chem_young}, and tracers of more chemically evolved gas like N$_2$H$^+$ and DCN, will provide insights into the origin of the streamer material \citep{Jaime2020_streamer, murillo_2022_streamer, tienhao_2023_streamer, taniguchi_2024_streamer}.
Chemically young regions are characterised by large abundances of atomic carbon in the gas phase (as carbon has not yet been mainly locked in CO), which is needed to produce carbon-chain molecules efficiently.

\section{Conclusions}
\label{sec_conclu}

We used high-sensitivity GBT observations of the pre-stellar core H-MM1 and its neighbourhood in the L1688 molecular cloud in Ophiuchus to study the multi-component structure of the gas towards the core and its immediate surroundings. Our results can be summarised as follows:

\begin{enumerate}
    \item Using a multi-component fit to the spectra in the line of sight, we separated the core and cloud in the line of sight, thereby allowing us to analyse the core material in more physical detail.
    
    \item We detect an extended subsonic region significantly larger than the previously observed coherent core: approximately 12 times larger in area, and approximately two times more massive.

    \item We find that the subsonic region is not continuous, but rather divided into two distinct velocity-coherent regions. One of these components, narrow-red, is spatially and kinematically linked to the dense core material. Therefore, we conclude that this component traces the core and the extended subsonic material connected to the core.

    \item Additionally, we detect a subsonic component, narrow-blue, which traces a more extended, lower-density region, with a kinetic temperature that is intermediate between that of typical dense cores and that of the ambient molecular cloud. There is a local peak observed in N(H$_2$) at the heart of this region, where \citet{ladj_2020_hgbs_oph_strc} report a starless bound core. Therefore, this extended component could be tracing material from a core in the early stages of formation.

    \item We find a narrow velocity structure towards the centre of the core from the extended subsonic region, with a smooth velocity gradient. We also observe a local increase in the velocity dispersion where this structure meets the inner part of the core. Therefore, we conclude that this feature likely shows the flow of material from the ambient cloud to the pre-stellar core H-MM1 via a streamer.

\end{enumerate}

The results from this work show the importance of a multi-component analysis and the usefulness of high-sensitivity observations. With our analysis, we have been able to separate multiple components in the line of sight and recover a much larger subsonic region than was previously observed. We have also found a previously undetected subsonic component likely associated with an early core. Moreover, we detect evidence of flow of gas from the ambient cloud to a pre-stellar core. Such accretion of chemically fresh gas onto dense cores during its evolution implies the need for a significant update to current physical and chemical models of core evolution.

\begin{acknowledgements}
C.W.L. acknowledges support from the Basic Science Research Program through the NRF funded by the Ministry of Education, Science and Technology (NRF- 2019R1A2C1010851) and by the Korea Astronomy and Space Science Institute grant funded by the Korean government (MSIT; project No. 2024-1-841-00).
\end{acknowledgements}

\bibliographystyle{aa}
\bibliography{biblio}

\begin{appendix}

\end{appendix}

\end{document}